# High fidelity of RecA-catalyzed recombination: a watchdog of genetic diversity

Dror Sagi, Tsvi Tlusty and Joel Stavans

Department of Physics of Complex Systems, The Weizmann Institute of Science, Rehovot 76100, Israel

**ABSTRACT**

**Homologous recombination plays a key role in generating genetic diversity, while maintaining protein functionality. The mechanisms by which RecA enables a single-stranded segment of DNA to recognize a homologous tract within a whole genome are poorly understood. The scale by which homology recognition takes place is of a few tens of base pairs, after which the quest for homology is over. To study the mechanism of homology recognition, RecA-promoted homologous recombination between short DNA oligomers with different degrees of heterology was studied *in vitro*, using fluorescence resonant energy transfer. RecA can detect single mismatches at the initial stages of recombination, and the efficiency of recombination is strongly dependent on the location and distribution of mismatches. Mismatches near the 5' end of the incoming strand have a minute effect, whereas mismatches near the 3' end hinder strand exchange dramatically. There is a characteristic DNA length above which the sensitivity to heterology decreases sharply. Experiments with competitor sequences with varying degrees of homology yield information about the process of homology search and synapse lifetime. The exquisite sensitivity to mismatches and the directionality in the exchange process support a mechanism for homology recognition that can be modeled as a kinetic proofreading cascade.**

## INTRODUCTION

Homologous recombination plays a key role in evolution and is an essential mechanism for the creation of genetic diversity (1,2). By reassembling sequences from homologous but not identical DNA molecules, as for example during horizontal transfer in unicellular organisms or crossover between chromosomes, recombination allows rapid acquisition of novel functions, driving adaptation and promoting speciation. Creation of new sequence combinations however has to be delicately balanced, to give the highest chance to the functional maintenance of coded proteins. Furthermore, the extent of recombination between related organisms, such as *Escherichia coli* and *Salmonella*, controls their genetic isolation and speciation (3). Therefore, inhibition of chromosomal gene transfer isolates related species.

In prokaryotes, recombination is carried out by RecA (4), a protein that also plays a fundamental role during the repair and bypass of DNA lesions, enabling the resumption of DNA replication at stalled replication forks. According to the prevailing view, the RecA-catalyzed recombination process proceeds along the following steps. First, a nucleoprotein complex is formed by the polymerization of RecA along a single-stranded DNA (ssDNA) substrate. ssDNA, which is accepted to be a prerequisite, is produced by the recombination-specific helicases RecBCD and RecG at double strand breaks (5), or as a byproduct of DNA repair processes at stalled replication forks (6). Next, there is a search for homology between the RecA-ssDNA nucleoprotein filament and double-stranded DNA (dsDNA). This search involves alignment of the nucleoprotein filament with a given tract along duplex DNA resulting in the formation of a three-stranded synaptic intermediate. Even in cases of heterology, formation of the synaptic intermediate involves unwinding of the duplex (7). Finally, a RecA-promoted strand exchange process between the single-stranded and double-stranded partners ensues, starting from the free 3' end of the invading duplex DNA (8).

The ability of RecA to speed up considerably the process of homology search within a huge excess of heterologous competitor sequences is an outstanding and unsolved problem. Whether resulting from horizontal gene transfer or from DNA repair processes, a given RecA-covered ssDNA segment locates an homologous partner sequence in a genome. Strikingly, *in vitro* evidence by Radding and coworkers indicates that RecA can identify its target within an excess of 200 000-fold heterologous substrate in $\sim$15 min. Homology must not be complete for recognition, and recombination can proceed with a limited degree of heterology between the invading strand and the target duplex DNA (1,9,10). For example, genomic rearrangements such as gene duplications can arise during the SOS response, as a result of recombination of partially homologous sequences such as the rearrangement hot spots *rhsA* and *rhsB* (11). In *E.coli*, the ability of RecA to tolerate heterologies during recombination is held in check by mismatch repair systems (MRS). There are however important situations in which MRS are either downregulated or inactivated. Bacteria experience feast and famine lifestyles, with the latter being more the rule rather than the exception. Under famine, when *E.coli* cells enter stationary phase, the MutHLS MRS is strongly downregulated (12–14). In natural *E.coli* isolates, 0.1–1.0% of cells are high mutators due to inactivation of *mut* genes. In these situations, the determinant mechanism to ensure approximate homology must be provided by the recombination process itself. Furthermore, in other bacteria such as *Streptococcus pneumoniae* or *Bacillus subtilis*, the MRS system is much less effective in regulating mismatches, and RecA-assisted recombination plays a decisive role in the discrimination of heterology (15,16). Despite the difference in the activity of the MRS between different bacteria types or mutants, recombination frequency decreases exponentially with increasing sequence divergence (3,15–17). Such a dependence cannot be explained by the stability of heteroduplex DNA, which varies linearly with the number of mismatches (18).

In contrast to these *in vivo* observations that suggest a high fidelity of the recombination process, numerous *in vitro* studies using long DNA substrates [thousands of basepairs (19–22)] have demonstrated that RecA is able to carry out strand exchange through large heterologies (20–22). It has been claimed that RecA does not discriminate significantly between perfect and imperfect matches of sequence, until the fraction of mismatches approaches 10% (9,19), and that RecA has an 'antiproofreading' activity with a sensitivity to mismatches below that expected in hybridization of naked DNA (23).

The need to provide a barrier against indiscriminate recombination in situations when MRS are not effective on one hand, and on the other the tolerance of large heterologies in recombination of long pieces of DNA hand, have led us to the hypothesis that the effects of discrimination by RecA may be decisive during the initial stages of recombination. To reexamine these effects, we have carried out a systematic, *in vitro* study of RecA-catalyzed recombination using fluorescence resonance energy transfer (FRET) techniques. The experimental scheme we used, shown in Figure 1, follows procedures introduced by previous workers (24–26). Linear dsDNA, labeled with a donor-acceptor pair, was combined with linear unlabeled ssDNA homologous to one of the strands of the duplex, in the presence of RecA. In this configuration, the donor-acceptor pair at the same side of the duplex, on the 5′ and 3′ ends of the complementary strands, yielded high transfer efficiency (TE) due to their close proximity. As a result of RecA-assisted strand exchange, the unlabeled strand displaces its homolog on the duplex, bringing the donor and acceptor far apart, thereby reducing the TE. In our scheme, the 5′ and 3′ labeled ends are staggered (see Materials and Methods). This design maximizes the TE and dynamic range of our measurements by avoiding the breakdown of the dipole–dipole approximation at short distances (27).

Our results show that RecA-mediated strand exchange is highly sensitive to the location and distribution of mismatches. A small fraction of properly located mismatches, which does not affect hybridization appreciably, can completely inhibit recombination. Moreover, even a single mismatch at the 'right' position can hamper strand exchange considerably. Accordingly, rejection of the incoming strand cannot be explained merely by the sensitivity of the annealing process to mismatches. Our results suggest a mechanism that uses ATP hydrolysis and directionality in the insertion of the invading strand to amplify the fidelity of homology search, similarly to kinetic proofreading (28,29). Experiments in the presence of heterologous competitor duplex DNA present evidence for a minimum efficient processing segment (MEPS), in which recombination is highly sensitive to mismatches. These results are consistent with an exponential dependence in DNA divergence for successful exchange, as was suggested *in vivo* (3). Thus, RecA does much more than mechanically exchanging strands and it can effectively compare DNA segments in its search for homology.

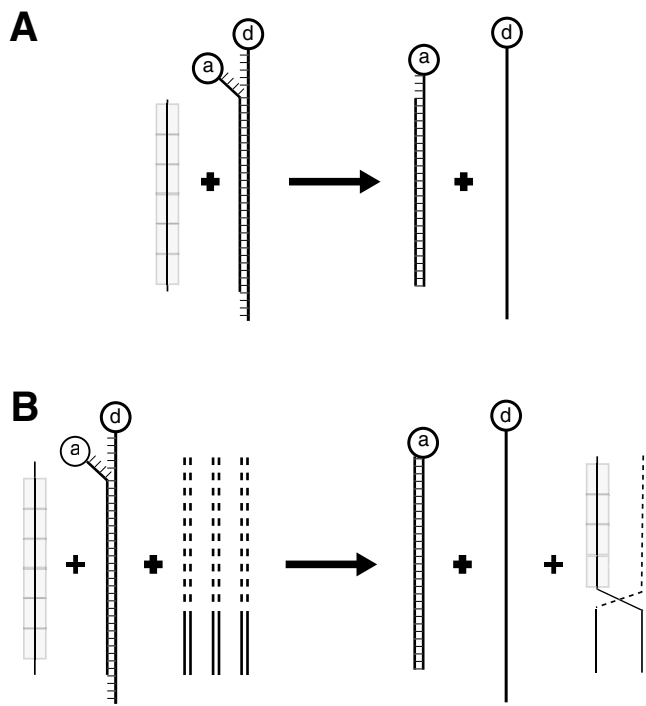

**Figure 1.** Experimental scheme to study RecA-induced recombination by FRET. (**A**) ssDNA–dsDNA recombination: a ssDNA substrate, henceforth referred to as the invading or incoming strand, covered with RecA (rectangles), is combined with dsDNA labeled at both strands on the same side of the duplex with a donor–acceptor pair, *d* and *a* respectively. Note the relative lengths of all oligomers (Materials and Methods). After strand exchange, no energy transfer takes place and only fluorescence from the donor is detected. (**B**) ssDNA–dsDNA recombination in the presence of competing duplexes with homology extending over part of their length.

## MATERIALS AND METHODS

### DNA oligonucleotides and RecA protein

ssDNA oligonucleotides labeled at their 3′ ends with tetramethylrhodamine (TAMRA) or at their 5′ ends with Cy5 fluorescent groups (Thermo Bioscience GmbH, Ulm, Germany), were HPLC and PAGE purified. Unlabeled oligonucleotides were synthesized and purified (PAGE) by the Synthesis Unit of the Biological Services Department of the Weizmann Institute of Science. The TAMRA and Cy5 labeled complementary strands were mixed in a 1:1 ratio in 10 mM HEPES–KOH (pH 7.5), 100 mM NaCl, 10 mM $MgCl_2$, and were hybridized by cooling from 90 to 20°C over 5 h. DNA sequences used in our experiments were all of natural origin and did not contain sequence motifs that could induce formation of non-canonical DNA structures. The labeled oligomeric sequences were designed with four mismatches (staggered ends) at the side of the duplex bearing the donor–acceptor pair, allowing to achieve near maximum TE by avoiding short-distance effects in resonant energy transfer (27). In addition, the acceptor-labeled strand was five bases longer from each side than the donor-labeled strand (Figure 1). Given that the persistence length of ssDNA is about 5 bp long (30), donor and acceptor were effectively dangling. To avoid spontaneous RecA-independent strand exchange, the sequence of the invading strands had also four mismatches at the complementary end opposing the fluorophore of the strand that switches partners. Exceptions were only sequences used in Figures 2A and 5, where the presence of the four mismatches did not affect our results. To increase the maximal rate of exchange as well as the dynamic range, concentrations of the invading ssDNAs were 2-fold than those of labeled duplexes. All ssDNA oligomers were checked for their secondary structure (31). We demanded less than eight successive intra-strand hydrogen bonds separated by fewer than four unpaired bases as by other workers (26). The sequences used in the experiments of Figure 3 are: (S1): 5′-ACTTCTACACTAGAAGGACAGTATTTGGTAT­CTGCGCTCTGCTGAAGCCAG-3′, (S2): 5′-GTTGCTCGT­TCGTTGCAACAAATTGATAAGCAATGCTTCTTGATA­ATGCAGTGAA-3′, (S3): 5′-CTGGCTTCAGCAGAGCGC­AGATACCAAATACTGTCCTTCTAGTGTAGTTCA-3′, (S4): 5′-TTCACTGCATTATCAAGAAGCATTGCTTATC­AATTTGTTGCAACGAACAGGTCTA-3′, (S5): 5′-ATTAT­TTCTCATTTTCCGCCAGCAGTCCACTTCGATTTAATT­CGTAAACA-3′, (S6): 5′-ACTTCTACACTAGAAGGACA­GTATTTGGTATCTGCGCTCTGCTGAAGCCAG-3′, (S7): 5′-TAGACCTGTTCGTTGCAACAAATTGATAAGCAAT­GCTTTTTTATAATGCAGTGAA-3′. Mismatches at a given position are obtained by changing G (C) to C (G) in the appropriate sequence. RecA protein was purchased from New England Biolabs.

### Sample preparation

FRET measurements were carried out in a reaction buffer containing 20 mM HEPES–KOH (pH 7.5), 2 mM ATP and 20 mM $MgCl_2$. Addition of 2 mM DTT and 100 μg/ml BSA had no measurable effect on our results. In the main protocol used in our assays, RecA, ssDNA and labelled dsDNA were mixed in reaction buffer to final concentrations of 2.5 μM, 200 and 100 nM respectively, incubated for 10 min at 37°C, and then injected into the measuring chamber. DNA concentrations are given as molar concentrations of oligonucleotide. To ensure that the amount of ATP is not appreciably depleted, the ATP concentration was ∼1000-fold that of RecA, following procedures by others (1,32,33). Under the conditions of our experiments (2 mM ATP and 10 min incubation times), inhibitory effects due to the accumulation of ADP are negligible (34). In assays probing recombination in the presence of competitor dsDNA of partial homology, the above protocol was modified by the addition of competitor duplexes together with the other reagents. To separate the timescale of filament fomation from other timescales such as those due to diffusional search and synapse lifetime, a second protocol in which filaments were preformed was used. In this protocol, ssDNA and RecA were incubated in reaction buffer at 37°C for 5 min, and then the labeled dsDNA and competitor were added and incubated for another 5 min. Both protocols yielded the same long time, steady state levels of strand exchange.

### FRET measurements and analysis

FRET measurements were performed using confocal fluorescence detection, and laser light illumination (Argon 514 nm line) in a home-built setup, as described in detail elsewhere (35). The intensities from both donor and acceptor channels were used to calculate TE as described previously. In our experiments, a decrease of donor fluorescence of 100 counts results in an increase in acceptor yield of 5–6 counts. Therefore, the factor γ, which measures the quantum yield, was estimated to be 0.05. Due to the configuration of the donor and acceptor in the labeled duplex, the donor's emission increases up to 9-fold in a strand displacement assay.

The fraction of exchange was extracted from donor and acceptor intensities as follows: first, the intensities from donor (d1) and acceptor (a1) channels were measured in a sample containing fully hybridized labeled duplexes. Next, the intensities from donor (d2) and acceptor (a2) channels were measured in a sample containing unhybridized, fully separated donor-labeled and acceptor-labeled ssDNA oligonucleotides, at the same concentration as in the duplex measurements. To ensure their separation during this measurement, the labeled ssDNAs were first hybridized with complementary unlabeled ssDNA partner strands. For a given RecA concentration, we assume that labeled ssDNA species are partitioned into doubly-labeled duplexes, or singly-labeled ones resulting from strand exchange. The fraction of doubly-labeled dsDNA having undergone strand exchange $x$ is then calculated from the equations:

$$d2 \cdot x + d1 \cdot (1-x) = d$$
$$a2 \cdot x + a1 \cdot (1-x) = a,$$

where $d$ and $a$ are the measured intensities of the donor and acceptor respectively. Each data point represents an average of three independent repetitions. Error bars are the standard deviations.

## RESULTS

### A small fraction of mismatches affects recombination significantly

To assess the efficiency of strand exchange as function of the degree of heterology, we combined dsDNA oligomers, doubly-labeled at one end with donor (5′ strand) and acceptor (3′ strand), with different ssDNA sequences shown in Figure 2A, at different RecA concentrations. Figure 2B shows the TE as a function of RecA concentration, and

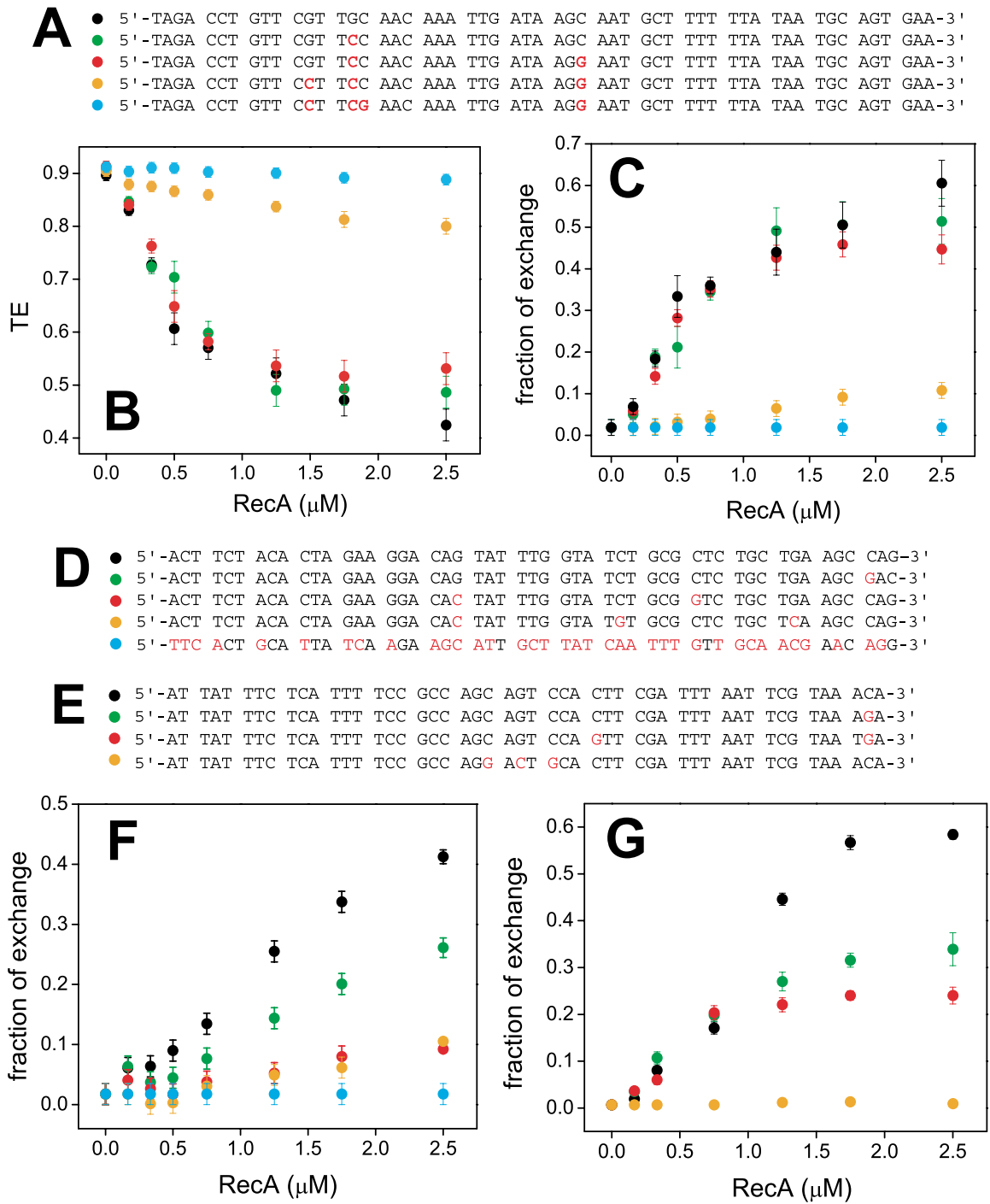

**Figure 2.** Effect of the number of mismatches on the efficiency of recombination. (**A**) Invading strands with an increasing number of mismatches (red), yield decreasing values of the transfer efficiency TE as a function of RecA concentration (**B**), and a lower corresponding fraction of oligomers having undergone strand exchange (**C**). The mismatches consist of G–C exchanges relative to the sequence labeled purely with black. Similar strand exchange experiments with two other different duplexes and corresponding invading ssDNA sequences with mismatches are shown in (**D** and **E**) and (**F** and **G**). All experiments were carried out with 2 mM ATP.

Figure 2C the corresponding values for the fraction of duplexes having undergone strand exchange. In these experiments, the A–T content and location was preserved, while mismatches were introduced by exchanging C–G. In experiments with full homology, TE decreases from a high value, 0.9, due to the close average proximity of donor and acceptor, to slighltly above 0.4 at 2.5 µM RecA. This corresponds to a fraction of exchange of ∼0.6, as shown in Figure 2C. While one or two mismatches did not change appreciably the efficiency of strand exchange, three mismatches lead however to large changes in TE and fraction of strand exchange, and four mismatches barely cause a reduction in TE and consequently, the fraction of exchange is very small (<0.02). Eight mismatches eliminated strand exchange completely (data not shown). To check that the results are sequence-independent, we carried out experiments with two completely different sequences (Figure 2D and E). We observed no RecA-independent strand exchange over a period of hours. In Figure 2F (corresponding to Figure 2D) one can see that 2 mismatches out of 51 bp reduce the fraction of strand exchange ∼4-fold, while Figure 2G (corresponding to Figure 2E) illustrates that three mismatches can reduce strand exchange to background levels. In conclusion, a small fraction of mismatches can significantly reduce the fraction of strand exchange.

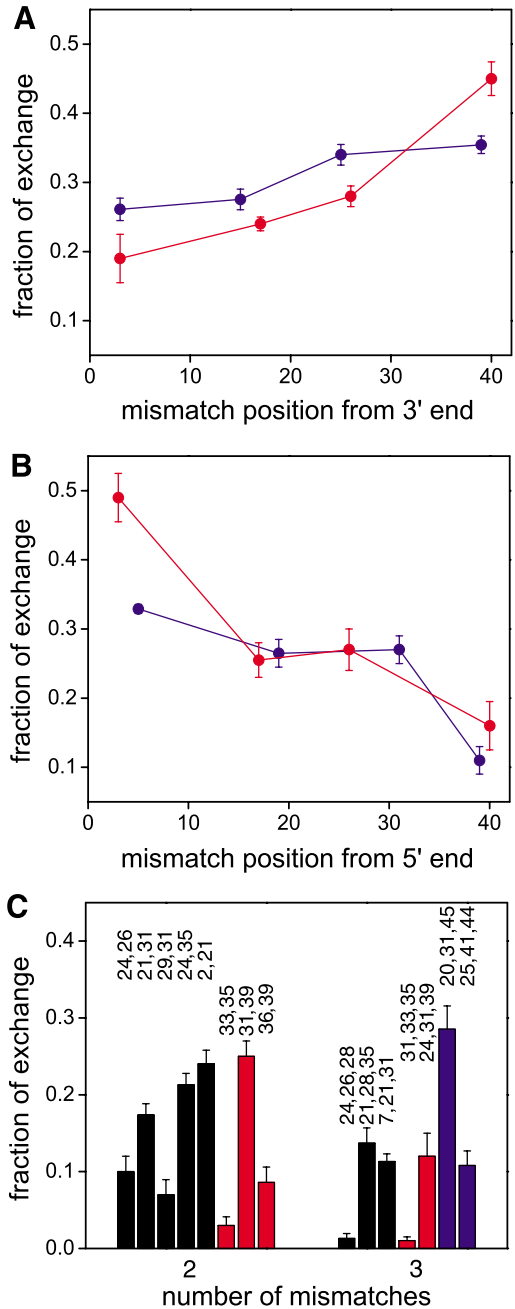

**Figure 3.** Effects of mismatch location and distribution on recombination. (**A**) Fraction of exchange as a function of distance of single mismatches from the 3′ end of the invading strand, for two different sequences S1 (blue) and S2 (red) (see Materials and Methods). The invading strand is complementary to the donor-labeled strand. For comparison purposes, the maximal values attained by the fraction of exchange in experiments with no mismatches were: 0.4 for S1, and 0.5 for S2. (**B**) As in B, but using sequences S3 (blue) and S4 (red) and distances are measured from the 5′ end of the invading strand. The invading strand is homologous to the donor-labeled strand. The maximal values attained by the fraction of exchange were 0.4 for S3 and 0.5 for S4. We stress that the x-axis in panels A and B represents the same position on the duplex (for details see text). (**C**) Fraction of exchange as a function of the number of mismatches, for different spatial distributions of mismatches, using sequences S5 (black), S6 (red) and S7 (blue) (see Materials and Methods). The numbers in the columns represent the mismatch distance from the 3′ end of the incoming strand. The maximal values attained by the fraction of exchange were 0.58 for S5, 0.4 for S6, and 0.6 for S7. Note that S5 corresponds to the sequences in Figure 2G and S6 corresponds to the sequences in Figure 2F.

### The initial stages of recombination are sensitive to mismatch location

The existing experimental evidence for the polarity of the strand exchange process (24) motivated us to examine the effects of mismatch location on recombination.

In Figure 3A we plot the fraction of exchange for different locations of single mismatches with respect to the 3′ end of the incoming strand, for two different sequences. In both cases the general trend is an increase of the efficiency of strand exchange as the mismatch moves away from the 3′ end. In these experiments, the invading strand is complementary to the donor-labeled strand in the duplex DNA. Thus, the x-axis represents the distance from the 5′ end of the donor-labeled strand. To verify that the observed trend is not due to the staggered ends near the fluorophores, we conducted experiments in which the invading strand is homologous to the donor-labeled strand. The results are plotted in Figure 3B as a function of the distance from the 5′ end of the incoming strand, which is also the distance from the 5′ end of the donor-labeled strand in the target duplex. Thus, for example, position 40 in panels A and B represents a unique location on the double-labeled duplex, 40 bp away from the 5′ end of the donor-labeled strand in the duplex DNA. With respect to the invading strand, this corresponds to position 40 relative to the 3′ end in panel A, and to position 16 relative to the 3′ end in panel B. The 3-fold difference in the fraction of exchange at this position between the two panels emphasizes the sensitivity to mismatch location and the directionality in the exchange process. To sum up, the data clearly shows that the extent of recombination increases when the distance of a mismatch from the 3′ end increases.

### Effects of mismatch distribution on the efficiency of strand exchange

We further enquired whether different spatial distributions of mismatches have an effect on the efficiency of strand exchange. In Figure 3C we plot the fraction of exchange for different spatial distributions of two and three mismatches. When mismatches are closely clustered their inhibitory effect on recombination is stronger than when they are far apart. The data in Figure 3C highlight that the fraction of exchange does not vary monotonically with the number of mismatches. For example, two clustered mismatches may lead to a smaller fraction of exchange than three mismatches that are far apart. To verify that the effect of clustered mismatches cannnot be attributed to a kinetic barrier, we also conducted experiments in which samples were incubated for 90 min. No change in the extent of exchange was observed. Thus our data reflect true steady state values.

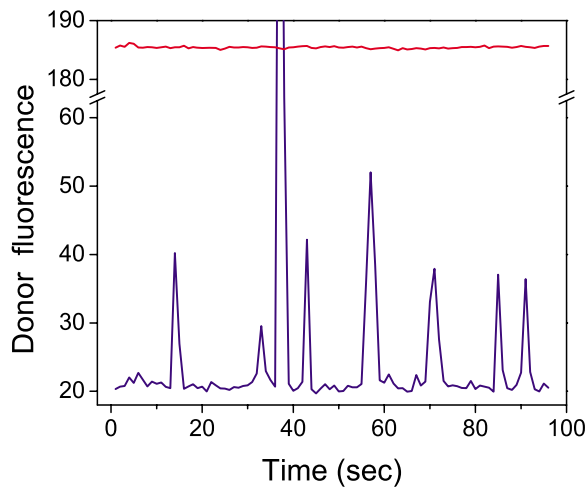

**Figure 4.** Effects of ATPγS in recombination: Fluorescence from the donor as a function of time in experiments with ATP (red) and ATPγS (blue).

### The effects of ATPγS

All the experiments described in previous sections were carried out using ATP. Given that the non-hydrolyzable form ATPγS has been used in the past (23,36), we tested for its effects. An interesting and unique feature of experiments carried out with ATPγS is the large extent of the fluctuations in the measured signals, either of the donor or the acceptor fluorescence. To illustrate this, we compare in Figure 4 temporal traces of donor fluorescence in the presence of either ATP or ATPγS. Each timepoint corresponds to a temporal average of the measured fluorescence over 1 s. Strikingly, the fluctuations of the fluorescence signal are much larger in the presence of ATPγS, and their typical timescale is ∼1 s. This behavior, characteristic of large, labeled objects crossing the focal region of the excitation beam, suggests that in the presence of both RecA and ATPγS, large aggregates of labeled DNA are formed.

Without ATP hydrolysis, RecA is unable to recycle between a state with high binding affinity to ssDNA, and a state with low binding affinity favoring depolymerization from the ssDNA substrate (37). Furthermore, it is known that ATPγS increases the affinity of RecA for ssDNA, and that a RecA monomer has two binding sites for DNA (38,39). Hence, in the presence of ATPγS, RecA-ssDNA filaments are stable and act as molecular stickers, linking together a number of labeled dsDNA molecules and forming a gel (40,41).

### Homology search in the presence of competing, partially-homologous sequences

The higher sensitivity of strand exchange to mismatches near the invading strand end, suggests that the stability of recombination intermediates formed after invasion of the 3′ end plays an important role during the search for homology. *In vivo* recombination measurements of phage-plasmid cointegrates in *E.coli* have revealed that for recombination to be efficient, the length over which homology extends must lie above a minimal value, the MEPS (42).

To shed light on the process of homology search and its characteristic timescales, test for the stability of recombination intermediates, and find out about RecA-related molecular mechanisms behind the MEPS, we conducted experiments in which RecA-ssDNA nucleoprotein complexes search for their fully homologous partners within labeled duplexes, in the presence of competing, unlabeled dsDNA oligomers of the same length (55 bp), but bearing homology only to tracts of different lengths $x$, including the 3′ end of the nucleoprotein complexes (Figure 1B). To increase the dynamic range of the exchange reaction, the invading strand was chosen to be fully complementary with its partner in the duplex, including the last 5 bp near its 5′ end. We note however that the bias that this choice provided in favor of exchange did not affect the interaction with the competitor sequences, as the 5′ end of the invading strand was nonetheless fully heterologous with the competitors (Figure 1B). In these assays, the concentration of the competitor duplexes was 10-fold that of the fully homologous labeled duplexes. Figure 5A shows the fraction of strand exchange $f$ as a function of time, under competition with different duplexes whose length of homology ranged from 0 to 35 bp. Typically the behavior was characterized by a fast rise at early times, and leveled off at long times. Both the rise time and the saturation value depend on the length of homology of the competitor DNA: when the length of homology was large, a competitor was more succesful at slowing down the rate at which the ssDNA finds its fully homologous partner, and furthermore, a larger fraction of the ssDNA was sequestered by the competitor at long times.

We have fit the curves in Figure 5A with an exponential dependence in order to extract a characteristic rise time $\tau$, for the fraction of exchange in the presence of the different competitors. We plot $\tau$ as a function of the length of homology of the competitor duplexes $x$ in Figure 5B. We also show the value of $\tau$ in the absence of competitor for comparison. There are two salient features in the behavior of $\tau$: a very weak increase for $x < 23$, followed by a higher rate of increase for $x > 25$. The behavior for $x > 23$ suggests that for a long enough stretch of homology, the RecA-ssDNA nucleoprotein filament is stabilized on the partially-homologous competitor DNA by Watson–Crick pairing, and that shorter stretches of homology are not stable enough and may exhibit a higher sensitivity to mismatches. The value $x$ ∼23 is similar to the *in vivo* estimate of the MEPS length (42). It is noteworthy that even for $x = 0$, i.e. no homology with the competitor duplexes, strand exchange is delayed by a significant amount (∼4 min). Given the experimental procedure followed, $\tau$ subsumes contributions due to the timescale associated with ssDNA-RecA filament formation by RecA nucleation and polymerization, the diffusional search of the filament for its duplex partner, synapse lifetime, and finally the time it takes for strand exchange. To factor out filament formation, we conducted experiments in which filaments were formed prior to adding the labeled duplex (see Materials and Methods). In the case of no competitor, these experiments yielded $\tau = 1.9 \pm 0.3$ min, instead of the $4.7 \pm 0.3$ min measured when following the first procedure.

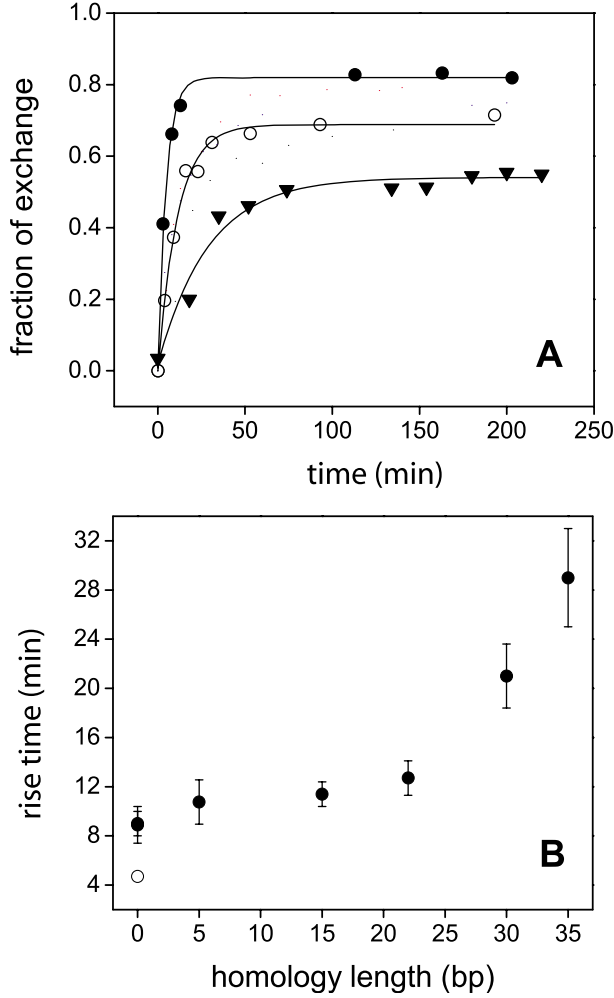

**Figure 5.** Effects of double-stranded competitor of partial homology on the strand exchange between fully homologous ssDNA and dsDNA. (**A**) Fraction of labeled duplexes having undergone strand exchange as a function time for competitor duplexes bearing homology for 0 (filled circles), 15 (open circles), and 35 (filled triangles) bp. Full lines are exponential fits to the data. (**B**) Rise time of the curves in (A) as function of the extent of homology of the competitor duplex. Also shown are rise times obtained from experiments using competitor duplexes bearing homology stretches 5, 22 and 30 bp long, not shown in panel (A) for clarity. In these reactions competitor concentration is 10-fold that of the labeled dsDNA. Empty circle: the ssDNA was preincubated with RecA for 5 min prior to mixing with dsDNA. Full circles: RecA, ssDNA and dsDNA were all added at the same time. In all experiments the RecA concentration was 2.5 µM.

The difference in $\tau$ between an experiment in which no competitor is present, to one in which competitor bearing no homology is present allowed us to separate the synapse lifetime from the process of strand exchange and estimate the former. This estimate is an important ingredient to understand the process of homology search within a cell. The diffusional timescale can be evaluated as the product of the Smoluchowski association rate (43) and the typical duplex concentration (100 nM) to be $\sim 10^{-2}$ s per collision between nucleoprotein filaments and target duplexes. Therefore, the 4 min difference in $\tau$ in the absence of competitor and in the presence of competitor bearing no homology with the labeled duplex is related to the synaptic lifetime. To estimate the lifetime $\tau_s$ of the synapse formed by the ssDNA-RecA filament with the competitor duplex, we express the rise time measured in the presence of competitor bearing no homology $\tau_{0c}$ ($x = 0$) as a sum of $\tau$ and $\tau_s$, in which the latter is weighted by the molar ratio $n_c/n_0$ between the labeled duplex that undergoes exchange and the competitor duplex:

$$\tau_{0c} = \tau + \frac{n_c}{n_0}\tau_s$$

Taking $\tau = 2$ min, $n_c/n_0 = 10$, and $\tau_{oc} = 6$ min results in $\tau_s = \sim 25$ s. Note, however, that the ratio $n_c/n_0$ is not constant during the reaction. During the exchange process the amount of labeled homologous duplex DNA decreases while the competitor remains unchanged. Therefore, $n_0$ decreases and $n_c$ remains unchanged, and the ratio $n_c/n_o$ increases. Thus, there is a correction factor of $\sim 2$ for $\tau_s$, which yields a synaptic lifetime, $\tau_s = 10$–15 s. We suspect that this characteristic time scale derives from the time it takes the RecA-coated oligomer to adjust its structure to that of the duplex DNA. The consistency between a time scale of 10 s for the synaptic lifetime with the time scale it takes to search in a whole genome, or when the competitor is in huge access (44) is discussed below.

## THEORETICAL MODEL

### Kinetic coarse-grained model

To elucidate the effects of mismatches, we describe the dynamics of RecA-catalyzed recombination as a three-stage process (Figure 6A). The initial state is a solution of dsDNA $a\bar{a}$ whose strands are homologous, together with the heterologous ssDNA $A$. We assume that that RecA has already polymerized along $A$. Within our simplified picture, the first stage is the physical contact of dsDNA, and the nucleoprotein complex that form the synapse $(a\bar{a})A$ (in our notation we put parentheses around the two strands connected by Watson–Crick bonds). The second stage is strand exchange whose product is the synapse $a(\bar{a}A)$. The third and last stage is the physical separation of the mismatched dsDNA $\bar{a}A$ from the ssDNA $a$. Each of the three stages has an inverse process. The chemical rate equations therefore take the form (Figure 5A):

$$\begin{aligned}
\frac{d}{dt}[A] &= \frac{d}{dt}[a\bar{a}] = -k_{on}[a\bar{a}][A] + k_{off}[(a\bar{a})A] \\
\frac{d}{dt}[(a\bar{a})A] &= k_{on}[a\bar{a}][A] - (k_{off} + k_f)[(a\bar{a})A] + k_b[a(\bar{a}A)] \\
\frac{d}{dt}[a(\bar{a}A)] &= k_{on}[\bar{a}A][a] + k_f[(a\bar{a})A] - (k_{off} + k_b)[a(\bar{a}A)] \\
\frac{d}{dt}[a] &= \frac{d}{dt}[\bar{a}A] = -k_{on}[\bar{a}A][a] + k_{off}[a(\bar{a}A)]
\end{aligned}$$

(1)

In this scheme, the potential impact of heterology is represented by the difference between $k_f$ and $k_b$ (Figure 6A). Neither the formation nor the disintegration of a synapse depend on the homology between participants within the framework of this model.

8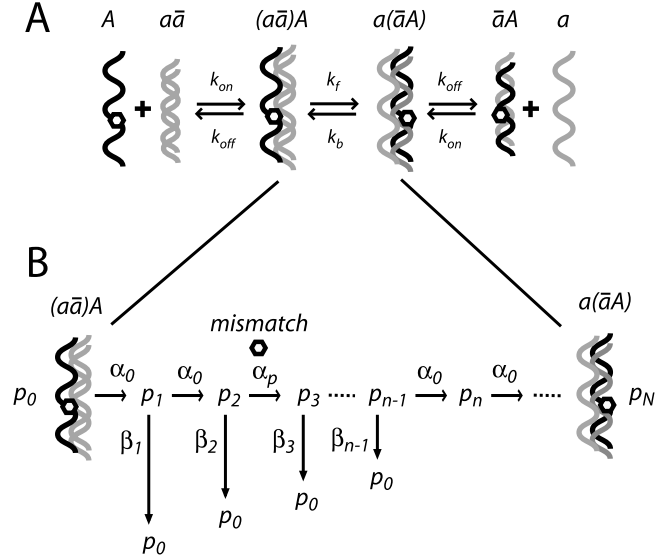

**Figure 6.** Theoretical scheme for homologous recombination. (**A**) Kinetics of the three-stage recombination process (for details see text). (**B**) Strand exchange via multi-stage kinetic proofreading: The exchange process advances step-by-step until it is either successfully completed or aborted. The probability that the synapse is at stage $j$ is $p_j$ (see text). Motivated by the measurements, the model assumes that synapse becomes more stable as the exchange advances. The corresponding disintegration rate $\beta_j$ therefore decreases from the 3′ end until it becomes negligible at the MEPS (stage $n$), as denoted by the shrinking arrows.

To compare with the measurements, we derive from Equation 1 the steady state fraction of recombined strands,

$$f = \frac{[a]}{n_0} = \frac{1}{1+\sqrt{k_b/k_f}} = \frac{1}{1+\sqrt{s}} \quad (2)$$

where $n_0$ is the total amount of the $a$ strand in ssDNA, dsDNA and synapses, $n_0 = [a] + [a\bar{a}] + [(a\bar{a})A] + [a(\bar{a}A)]$ and $s = k_b/k_f$ is a measure for asymmetry. With no mismatches, $k_f = k_b$ and the fraction of exchange is symmetric, $f = 1/2$. The deviation from symmetry indicates slowdown of the exchange by mismatches. The high fraction of exchange in our measurements indicates that most strands are not forming part of synapses. Hence the physical formation of a synapse and its disintegration are much faster than strand exchange. In terms of rates, this experimental regime implies $k_{on} n_0 \ll k_{off}$.

The kinetic equations, Equation 1, portray the exchange process as a single step characterized by a single effective rate $k_f$ or $k_b$. In the following, we replace this simplified picture by a description of the exchange in terms of a sequential multi-step process that accounts for the effects of mismatch position and distribution.

### Sequential check and exchange—kinetic proofreading cascade

Our measurements demonstrated high sensitivity of RecA-mediated recombination to mismatches, their location and distribution. Amplification of the sensitivity of molecular recognition at the expense of energy is the hallmark of a kinetic proofreading mechanism. Energy is consumed by irreversible recognition steps that increase the sensitivity in the same manner as the stages of a refinery increase the purity of an enriched liquid (45). Furthermore, measurements in the presence of competing, partially homologous, strands indicated that the exchange time scales linearly with the strand length. This motivated us to treat the exchange process in terms of a sequential step-by-step dynamics. Exchange initiates by the invasion of the dsDNA by the RecA-covered ssDNA near the 3′ end (24) and advances linearly, in a process that involves (i) the alignment of both participant strands, (ii) stretching of the dsDNA to fit the longer pitch of the RecA polymer, (iii) homology check and (iv) exchange. The precise details of the molecular mechanism that drives this process are yet to be elucidated. We focus on the generic properties of such a sequential process that shares many similarities with a cascaded kinetic proofreading scheme (28, 46, 47). An essential feature of the model is the directionality of the search and exchange process. With directional dynamics, the exchange time scales linearly with the strand length $L$, while diffusive dynamics lead to an exchange time that scales like $L^2$. The directional, irreversible advance requires energy consumption in the form of ATP. The dynamics is that of an irreversible multi-step process (Figure 6B): The synapse has a probability $p_j$ to be at step $j$, and it can then advance either forward to the next state at a rate $\alpha_j$, or backwards all the way to the first stage, at a rate $\beta_j$

$$\frac{d}{dt}p_j = -\alpha_j p_j + \alpha_{j-1} p_{j-1} - \beta_j p_j. \quad (3)$$

Balancing the influx and outflux from each step we find that the ratio $1/s$ of final to initial step probabilities at steady state is:

$$\frac{1}{s} = \frac{p_N}{p_0} = \prod_{j=1}^{N} \frac{\alpha_{j-1}}{\alpha_j + \beta_j}. \quad (4)$$

It is expected that the synapse becomes more stable as the exchange advances. We therefore assume that the corresponding disintegration rate $\beta_j$ decreases from the 3′ end until it becomes negligible at the MEPS. The forward rate on the other hand remains roughly constant $\alpha_0$ for all steps that involve homologous triplets. The impact of a mismatch is to delay the forward advancement by reducing the forward rate to $\alpha_p \ll \alpha_0$. It follows from Equation 4 that the attenuation by a mismatch at site $j$ is

$$m_j = \frac{s_0}{s} = \frac{1+\beta_{j-1}/\alpha_0}{1+\beta_{j-1}/\alpha_p}. \quad (5)$$

It is noteworthy that a mismatch has a minor effect beyond the MEPS when the disintegration rate $\beta_j$ virtually vanishes. On the other hand, when a mismatch is present near the 3′ end its effect is dramatic. This sensitivity to the location of the mismatch results from the irreversibility of the scheme (Figure 6B). In precise analogy to kinetic proofreading (45–47), the energy-driven repetition of irreversible recognition steps amplifies even minute sequence heterology.

In the case of well-separated mismatches, the combined effect is to reduce the exchange by a factor that is the product of their individual attenuation factors given by

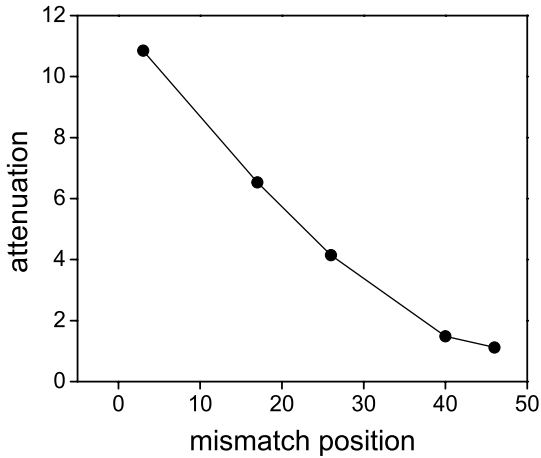

**Figure 7.** Normalized values of the attenuation factor s derived from the steady state values of the fraction of strand exchange $f$ in Figure 3 and Equation 2, in recombination experiments with single mismatches at different positions relative to the 3' end of the invading strand. The values of s were normalized by the value $s_0$ obtained with full homology. The line is a guide to the eye.

Equation 5, $m_T = \prod_{j=1}^{N} m_j$. On the other hand, adjacent mismatches reduce the forward rate much below $\alpha_p$, thus destabilizing the synapse more than if mismatches were far apart.

Taking into account point mutations at a probability per site $\mu$, the average total attenuation is given by:

$$\langle m_T \rangle = \prod_{j=1}^{N} ((1-\mu) + \mu m_j) \approx \exp\left(-\mu \sum_{j=1}^{N}(1-m_j)\right) \approx \exp(-\mu L), \quad (6)$$

where we employed the approximation that within a MEPS, $L$, the individual attenuations are strong $m_j \ll 1$. The exponential sensitivity of the recombination efficiency on both the length of the MEPS and the mismatch probability is a direct consequence of the sequential nature of the exchange mechanism. Such sensitivity may set a well-defined genetic barrier between species and individuals.

### Consistency with experimental observations

In spite of its simplicity, the kinetic model yields a specific prediction for the dependence of the fraction of strand exchange $f$ as a function of the asymmetry parameter $s = k_b/k_f$ (Equation 2). It is interesting to compare the consistency of this prediction with our observations of the MEPS. We show in Figure 7 the calculated values of $s$ (Equation 2) in the case of single mismatches at different positions from the 3' end, normalized by the value $s_0$ obtained when no mismatch is present $\tilde{s} = s/s_0$. The values of $\tilde{s}$, which are actually the attenuation factors (Equation 5, decrease monotonously with the distance of the mismatch from the 3' end, and converge to 1. At position 25, the effect of mismatch has decreased to half its maximum. Thus the scale over which the decay takes place, roughly 25 bp, which is expected to be roughly of the same size as the MEPS, coincides with the measured value in our competition assays.

Finally, we note that the kinetic proofreading cascade scheme can also account for the lower extent of recombination observed when the G–C content in the participant strands is increased (24). A G–C Watson–Crick pair may reduce the forward rate $\alpha$ relative to an A–T pair, thus having the same qualitative effect as introducing mismatches.

## DISCUSSION

Two key features revealed by the present study are an exquisite sensitivity during the initial stages of strand exchange process manifested as the ability to discriminate single mismatches, and the strong dependence of the recombination process on the distribution and location of mismatches. These findings highlight the profound differences between the discrimination of heterology by RecA and hybridization, in which discrimination is dominated by the fidelity of Watson–Crick pairing. Thus, the function of RecA is not limited to catalyzing a purely mechanical exchange of strands. The sensitivity to the location of mismatches is best illustrated by their strong effect on the efficiency of recombination when located near the 3' end of the invading strand, as opposed to their negligible effects when they are located near the 5' end, providing direct evidence for the directionality of the strand invasion process. This behavior is consistent with the polymerization and depolymerization of RecA from the 5' to the 3' end (48), and suggests a sequential mechanism for homology recognition.

We note that while RecA has been reported to decrease the fidelity of homology search below that of hybridization (23,49), these studies only probed the formation of the synaptic complex rather than strand exchange. In addition, these studies used ATPγS and ADP rather than ATP. It is conceivable that in the absence of ATP hydrolysis the pairing reaction has a low sensitivity to mismatches.

Our experiments with competitor DNA sequences of prescribed and varying homology provide a molecular basis for the *in vivo* observation of a MEPS (42). *In vitro* this is manifested by the near insensitivity of the sequestering time on the degree of homology of competitor sequences, when the length of homology tracts lies below ~25 bp, and its observed increase above this length. In this latter case hybridization stabilizes recombination intermediates, and the sensitivity to mismatches decreases sharply. Consistent with these results is the decrease of the attenuation factor to single mismatches with the distance of the mismatch from the 3' end.

The competition experiments also allowed us to estimate the delay in fully homologous recombination deriving from the presence of competitor sequences bearing no homology with the recombining participants. Under the conditions of our experiments, using 55 bp long DNA oligomers our estimate of this delay is ~10 s. We conjecture that this delay results among other factors from the necessity to match the different geometries of the coated ssDNA and the duplex target, and test for homology. Our estimate is seemingly inconsistent by orders of magnitude with other *in vitro* experiments, in which recombination of a single-stranded ~10 kbp plasmid was monitored in the presence of 200 000 excess competitor (44). In these experiments the reaction was 55% complete after only 10 min. Indeed, assuming the delay scales with the concentration of competitor sequences, one would expect from our measurements a delay of $\sim 10^6$ s for a 100 000 excess of competitor. Our estimate of the synapse lifetime and the results of Honigberg *et al.* (44) can be reconciled by postulating that the search of homology in the case of long substrates is carried out in parallel, with the invading strand targeting numerous disjoint MEPS segments on the target DNA simultaneously, a proposition that has been advanced before (40). The search for homology has been addressed before in experiments in which ATPγS was used (50). The use of this cofactor introduces irreversibility since the dissociation of the synaptic complex is strongly reduced (40). Therefore pairing and not strand exchange was monitored.

The sensitivity to the location and distribution of mismatches as well as the degree of discrimination achieved by RecA during recombination, are considerably higher than that afforded by Watson–Crick pairing between the invading strand and its complementary strand in the target duplex. Two characteristics are critical to achieve this high amplification in discrimination: the directionality of the recombination process, and energy expenditure. These characteristics motivated the use of kinetic proofreading ideas (28,29) to build a theoretical scheme consisting of a cascade of irreversible events (45–47). The calculations show that these features enable an exponential decrease with DNA divergence between partners, in agreement with *in vivo* experiments (3,15). A scheme in which strand exchange is a sequential, unidirectional process is expected to be highly sensitive to mismatches: a given mismatch can only be reached if all previous mismatches have been overcome. This however is expected to apply as long as the extent of strand displacement is below the length of a MEPS. Above this length the sensitivity to mismatches decreases rapidly, and the search for homology is over. We note in passing that the RecA-mediated aggregation of DNA strands observed when using ATPγS did not allow us to establish the precise role that ATP hydrolysis plays during the processes of homology recognition and strand exchange.

A necessary condition for speciation to arise is the establishment of genetic barriers between closely related organisms. Bacteria, as opposed to eukaryotes, may be highly promiscuous in the transfer of genetic material through transformation, transduction and conjugation processes. While the rate of homologous recombination may be lower than that of mutations, the available philogenetic range makes this transfer a potent mechanism for adaptative evolution. The exquisite sensitivity to single mismatches exhibited by RecA-induced recombination as revealed by the present study suggests that homologous recombination alone can play a key role in maintaining genetic isolation in bacteria, particularly in situations in which MRS are either downregulated as during stationary phase in enterobacteria, or when MRS do not play a dominant role. An interesting issue remaining for future studies is whether the high degree of discrimination exhibited by RecA during recombination was preserved during evolution and is exhibited in eukaryotic recombinases as well.

## REFERENCES

1. Bucka,A. and Stasiak,A. (2001) RecA-mediated strand exchange traverses substitutional heterologies more easily than deletions or insertions. *Nucleic Acids Res.*, **29**, 2464–2470.
2. Griffiths,A.J.F., Miller,J.H., Suzuki,D.T., Lewontin,R.C. and Gelbart,W.M. (1996) *An Introduction to Genetic Analysis*. W.H. Freeman and Company, New York, USA.
3. Vulic,M., Dionisio,F., Taddei,F. and Radman,M. (1997) Molecular keys to speciation: DNA polymorphism and the control of genetic exchange in enterobacteria. *Proc. Natl Acad. Sci. USA*, **94**, 9763–9767.
4. Cox,M.M. (1999) Recombinational DNA repair in bacteria and the RecA protein. *Prog. Nucleic Acid Res. Mol. Biol.*, **63**, 311–366.
5. Kowalczykowski,S.C., Dixon,D.A., Eggleston,A.K., Lauder,S.D. and Rehrauer,W.M. (1994) Biochemistry of homologous recombination in *Escherichia coli*. *Microbiol. Rev.*, **58**, 401–465.
6. Kowalczykowski,S.C. (2000) Initiation of genetic recombination and recombination-dependent replication. *Trends Biochem. Sci.*, **25**, 156–165.
7. Rould,E., Muniyappa,K. and Radding,C.M. (1992) Unwinding of heterologous DNA by Reca protein during the search for homologous sequences. *J. Mol. Biol.*, **226**, 127–139.
8. Friedman-Ohana,R. and Cohen,A. (1998) Heteroduplex joint formation in *Escherichia coli* recombination is initiated by pairing of a 3′-ending strand. *Proc. Natl Acad. Sci. USA*, **95**, 6909–6914.
9. Bazemore,L.R., FoltaStogniew,E., Takahashi,M. and Radding,C.M. (1997) RecA tests homology at both pairing and strand exchange. *Proc. Natl Acad. Sci. USA*, **94**, 11863–11868.
10. Rao,B.J. and Radding,C.M. (1994) Formation of base triplets by non-Watson–Crick bonds mediates homologous recognition in RecA recombination filaments. *Proc. Natl Acad. Sci. USA*, **91**, 6161–6165.
11. Dimpfl,J. and Echols,H. (1989) Duplication mutation as an SOS response in *Escherichia coli*—enhanced duplication formation by a constitutively activated RecA. *Genetics*, **123**, 255–260.
12. Rayssiguier,C., Thaler,D.S. and Radman,M. (1989) The barrier to recombination between *Escherichia coli* and *Salmonella typhimurium* Is disrupted in mismatch-repair mutants. *Nature*, **342**, 396–401.
13. Bregeon,D., Matic,I., Radman,M. and Taddei,F. (1999) Inefficient mismatch repair: genetic defects and down regulation. *J. Genet.*, **78**, 21–28.
14. Feng,G., Tsui,H.C.T. and Winkler,M.E. (1996) Depletion of the cellular amounts of the MutS and MutH methyl-directed mismatch repair proteins in stationary-phase *Escherichia coli* K-12 cells. *J. Bacteriol.*, **178**, 2388–2396.
15. Majewski,J., Zawadzki,P., Pickerill,P., Cohan,F.M. and Dowson,C.G. (2000) Barriers to genetic exchange between bacterial species: *Streptococcus pneumoniae* transformation. *J. Bacteriol.*, **182**, 1016–1023.
16. Majewski,J. and Cohan,F.M. (1999) DNA sequence similarity requirements for interspecific recombination in bacillus. *Genetics*, **153**, 1525–1533.
17. Delmas,S. and Matic,I. (2005) Cellular response to horizontally transferred DNA in *Escherichia coli* is tuned by DNA repair systems. *DNA Repair (Amst)*, **4**, 221–229.
18. Springer,M.S., Davidson,E.H. and Britten,R.J. (1992) Calculation of sequence divergence from the thermal stability of DNA heteroduplexes. *J. Mol. Evol.*, **34**, 379–382.
19. DasGupta,C. and Radding,C.M. (1982) Polar branch migration promoted by recA protein: effect of mismatched base pairs. *Proc. Natl Acad. Sci. USA*, **79**, 762–766.
20. West,S.C., Cassuto,E. and Howard-Flanders,P. (1981) Heteroduplex formation by RecA protein: polarity of strand exchanges. *Proc. Natl Acad. Sci. USA*, **78**, 6149–6153.
21. Bianchi,M.E. and Radding,C.M. (1983) Insertions, deletions and mismatches in heteroduplex DNA made by recA protein. *Cell*, **35**, 511–520.
22. Morel,P., Stasiak,A., Ehrlich,S.D. and Cassuto,E. (1994) Effect of length and location of heterologous sequences on RecA-mediated strand exchange. *J. Biol. Chem.*, **269**, 19830–19835.
23. Malkov,V.A., Sastry,L. and Camerini-Otero,R.D. (1997) RecA protein assisted selection reveals a low fidelity of recognition of homology in a duplex DNA by an oligonucleotide. *J. Mol. Biol.*, **271**, 168–177.

24. Gupta,R.C., Golub,E.I., Wold,M.S. and Radding,C.M. (1998) Polarity of DNA strand exchange promoted by recombination proteins of the RecA family. *Proc. Natl Acad. Sci. USA*, **95**, 9843–9848.
25. Gumbs,O.H. and Shaner,S.L. (1998) Three mechanistic steps detected by FRET after presynaptic filament formation in homologous recombination. ATP hydrolysis required for release of oligonucleotide heteroduplex product from RecA. *Biochemistry*, **37**, 11692–11706.
26. Folta-Stogniew,E., O'Malley,S., Gupta,R., Anderson,K.S. and Radding,C.M. (2004) Exchange of DNA base pairs that coincides with recognition of homology promoted by *E. coli* RecA protein. *Mol. Cell*, **15**, 965–975.
27. Schuler,B., Lipman,E.A., Steinbach,P.J., Kumke,M. and Eaton,W.A. (2005) Polyproline and the 'spectroscopic ruler' revisited with single-molecule fluorescence. *Proc. Natl Acad. Sci. USA*, **102**, 2754–2759.
28. Hopfield,J.J. (1974) Kinetic proofreading: a new mechanism for reducing errors in biosynthetic processes requiring high specificity. *Proc. Natl Acad. Sci. USA*, **71**, 4135–4139.
29. Ninio,J. (1975) Kinetic amplification of enzyme discrimination. *Biochimie*, **57**, 587–595.
30. Allemand,J.F., Bensimon,D. and Croquette,V. (2003) Stretching DNA and RNA to probe their interactions with proteins. *Curr. Opin. Struct. Biol.*, **13**, 266–274.
31. Zuker,M. (2003) Mfold web server for nucleic acid folding and hybridization prediction. *Nucleic Acids Res.*, **31**, 3406–3415.
32. Holmes,V.F., Benjamin,K.R., Crisona,N.J. and Cozzarelli,N.R. (2001) Bypass of heterology during strand transfer by Saccharomyces cerevisiae Rad51 protein. *Nucleic Acids Res.*, **29**, 5052–5057.
33. Noirot,P., Gupta,R.C., Radding,C.M. and Kolodner,R.D. (2003) Hallmarks of homology recognition by RecA-like recombinases are exhibited by the unrelated *Escherichia coli* RecT protein. *EMBO J.*, **22**, 324–334.
34. Lee,J.W. and Cox,M.M. (1990) Inhibition of recA protein promoted ATP hydrolysis. 1. ATP gamma S and ADP are antagonistic inhibitors. *Biochemistry*, **29**, 7666–7676.
35. Sagi,D., Friedman,N., Vorgias,C., Oppenheim,A.B. and Stavans,J. (2004) Modulation of DNA conformations through the formation of alternative high-order HU-DNA complexes. *J. Mol. Biol.*, **341**, 419–428.
36. Kowalczykowski,S.C. and Krupp,R.A. (1995) DNA-strand exchange promoted by RecA protein in the absence of ATP—implications for the mechanism of energy transduction in protein-promoted nucleic-acid transactions. *Proc. Natl Acad. Sci. USA*, **92**, 3478–3482.
37. Rosselli,W. and Stasiak,A. (1990) Energetics of RecA-mediated recombination reactions—without ATP hydrolysis RecA can mediate polar strand exchange but is unable to recycle. *J. Mol. Biol.*, **216**, 335–352.
38. Muller,B., Koller,T. and Stasiak,A. (1990) Characterization of the DNA-binding activity of stable RecA-DNA complexes—interaction between the 2 DNA-binding sites within RecA helical filaments. *J. Mol. Biol.*, **212**, 97–112.
39. Zaitsev,E.N. and Kowalczykowski,S.C. (1999) The simultaneous binding of two double-stranded DNA molecules by *Escherichia coli* RecA protein. *J. Mol. Biol.*, **287**, 21–31.
40. Dutreix,M., Fulconis,R. and Viovy,J.-L. (2003) The search for homology: a paradigm for molecular interactions? *ComPlexUs*, **1**, 89–99.
41. Tsang,S.S., Chow,S.A. and Radding,C.M. (1985) Networks of DNA and RecA protein are intermediates in homologous pairing. *Biochemistry*, **24**, 3226–3232.
42. Shen,P. and Huang,H.V. (1986) Homologous recombination in *Escherichia coli*—dependence on substrate length and homology. *Genetics*, **112**, 441–457.
43. Halford,S.E. and Marko,J.F. (2004) How do site-specific DNA-binding proteins find their targets? *Nucleic Acids Res.*, **32**, 3040–3052.
44. Honigberg,S.M., Rao,B.J. and Radding,C.M. (1986) Ability of RecA protein to promote a search for rare sequences in duplex DNA. *Proc. Natl Acad. Sci. USA*, **83**, 9586–9590.
45. Bar-Ziv,R., Tlusty,T. and Libchaber,A. (2002) Protein-DNA computation by stochastic assembly cascade. *Proc. Natl Acad. Sci. USA*, **99**, 11589–11592.
46. McKeithan,T.W. (1995) Kinetic proofreading in T-cell receptor signal-transduction. *Proc. Natl Acad. Sci. USA*, **92**, 5042–5046.
47. Tlusty,T., Bar-Ziv,R. and Libchaber,A. (2004) High-fidelity DNA sensing by protein binding fluctuations. *Phys. Rev. Lett.*, **93**, 258103.
48. Cox,M.M. (2003) The bacterial RecA protein as a motor protein. *Annu. Rev. Microbiol.*, **57**, 551–577.
49. Malkov,V.A. and Camerini-Otero,R.D. (1998) Dissociation kinetics of RecA protein-three-stranded DNA complexes reveals a low fidelity of RecA-assisted recognition of homology. *J. Mol. Biol.*, **278**, 317–330.
50. Yancey-Wrona,J.E. and Camerini-Otero,R.D. (1995) The search for DNA homology does not limit stable homologous pairing promoted by RecA protein. *Curr. Biol.*, **5**, 1149–1158.